\documentstyle[preprint,eqsecnum,aps]{revtex}

\begin{document}
\tightenlines
 \draft
\title{TWIST MODE IN SPHERICAL ALKALI METAL CLUSTERS}
\author{V.O. NESTERENKO$^{1,2}$, J.R. MARINELLI$^1$, F.F. de SOUZA
CRUZ$^{1}$,}
\address{W. KLEINIG$^{2,3}$, and P.-G. REINHARD$^4$}
\date{\today}
\address{$^{1}$Departamento de Fisica, Universidade Federal de Santa
Catarina, Florian$\acute o$polis, SC, 88040-900, Brasil}
\address{$^{2}$Bogoliubov Laboratory of Theoretical Physics,
Joint Institute for Nuclear Research, Dubna, Moscow region, 141980, Russia}
\address{$^{3}$Technische Universit$\ddot a$t Dresden,
Institut f$\ddot u$r Analysis, Dresden, D-01062, Germany}
\address{$^{4}$Institut f\"ur Theoretische Physik,
Universit\"at Erlangen, D-91058, Erlangen, Germany}
\maketitle

\begin{abstract}
A remarkable orbital quadrupole magnetic resonance, so-called twist mode, is
predicted in alkali metal clusters where it is represented by $I^{\pi}=2^-$
low-energy excitations of valence electrons with strong M2 transitions to
the ground state. We treat the twist by both macroscopic and microscopic
ways. In the latter case, the shell structure of clusters is fully
exploited, which is crucial for the considered size region ($8\le N_e\le
1314$). The energy-weighted sum rule is derived for the pseudo-Hamiltonian.
In medium and heavy spherical clusters the twist dominates over
its spin-dipole counterpart and becomes the most strong multipole magnetic mode.
\end{abstract}


\arraycolsep1.5pt







\narrowtext

\preprint{HEP/123-qed}

\narrowtext

Orbital magnetism in atomic clusters is a subject of special interest.
Clusters may contain many atoms and, therefore, single-particle orbital
moments of valence electrons can reach very big values. This results in huge
orbital effects, for example, in strong orbital magnetic resonances. These
resonances are of a general character and exist in different finite Fermi
systems (nuclei, atomic clusters, etc). M1 scissor mode in deformed systems
\cite{Iud}-\cite{Ne-PRL} and M2 twist mode in systems of arbitrary shape
\cite{Lamb}-\cite{Bast} are most famous examples. They have been observed in
atomic nuclei (see, e.g., \cite{Rich,Pon}) but not yet in clusters. The
scissors mode has been already predicted in clusters\cite{LS-ZPD,Ne-PRL}. In
the present paper we will discuss properties of the twist mode
in this system.

By definition, the twist is the quadrupole torsional vibration mode of an
elastic globe\cite{Lamb}. It is generated by the operator ${\hat T}%
=e^{-i\alpha z l_z}=e^{\alpha {\vec u} \cdot {\vec \bigtriangledown}}$ with
the velocity field ${\vec u}=(yz,-xz,0)$\cite{Lamb,HE}. The mode is viewed
macroscopically (see Fig. 1) as small-amplitude rotation-like
oscillations of different layers of a system against each other with a
rotational angle proportional to $z$ (projection to the axis of rotation).
The restoring force of the twist is determined by the quadrupole distortions
of the Fermi surface in the {\it momentum} space. So, the twist represents
transverse magnetic quadrupole oscillations of an {\it elastic} medium,
provided by variations of the kinetic-energy density. The twist is a general
feature of any 3-dimensional finite Fermi system which demonstrates an
elastic behavior. Atomic nuclei\cite{HE}-\cite{Pon} and clusters\cite{Bast}
are most typical examples. Unlike the M1 scissor mode which has the similar
quantum origin but can exist only in deformed systems, the twist manifests
itself in Fermi systems of any shape, spherical and deformed.

Expressions for the twist energy and M2 strength, obtained within the
elastodynamical models\cite{HE,Bast_n} for atomic nuclei, can be
reformulated for clusters as
\begin{equation}  \label{eq:elast}
\omega = 17 eV \AA^2 r_s^{-2} N_e^{-1/3}, \quad B(M2)=0.52 r_s^2 N_e^2
\mu_b^2
\end{equation}
where B(M2) is probability of M2 transition from the ground state to $2^-$
twist state, $r_s$ is the Wigner-Seitz radius, and $N_e$ is the number of
valence electrons. We will show that
expressions (\ref{eq:elast}) provide good {\ qualitative} agreement with
microscopic results.
However, elastodynamical models do not access the shell structure of
nanoparticles and so cannot clarify the microscopic origin of the twist.
Moreover, these models are questionable for small systems. In this paper we
will present, for the first time, a microscopic analysis of the twist, fully
embracing shell effects.
Both small and heavy clusters will be covered.

The operator for $M2$ transition\cite{BW}, ${\hat{F}}(M2,\mu)=
\mu _b\sqrt{10}r[g_s\{Y_1{\hat{s}}\}_{2\mu}+\frac 23g_l\{Y_1{\hat{l}}%
\}_{2\mu}], $ is a sum of spin and orbital components with $g_s=2$ and $%
g_l=1 $. The external field generating the twist, $zl_z\propto r(Y_{10}l_z)$%
, is a part of the orbital component with $\mu=0$. So, it is natural to
consider the twist as a part of the orbital M2 resonance. Since both are of
a similar nature, we will call the whole orbital M2 resonance also as a
twist.

In spherical systems, only the spin-dipole channel delivers the residual
interactions for $2^-$ excitations. Investigations \cite{Schw,KINK} have
shown that twist in spherical atomic nuclei only weakly depends on this
residual interaction and the remaining dependence is mainly caused by the
spin-orbit coupling. The influence of the residual interaction should then
be even smaller for metal clusters where the spin-orbit coupling is
negligible. Thus we can deal with a simple particle-hole (1p-1h) picture.
Also, we will
confine our consideration to the spherical jellium model which is compulsory
for analysis of large systems. These two approximations are appropriate
for the present purpose of a first survey of the twist mode. At the side of
the mean field Hamiltonian, we take into account local as well as non-local
effects which are caused by the ionic pseudo-potential.
For Na and K clusters we calculate the Kohn-Sham single-particle scheme within
the approach\cite{SAJM} which properly treats
the local effects. For Li clusters, the pseudo-Hamiltonian\cite{PH}
\begin{equation}
H_0=-\frac{\hbar^2}{2m_e}\vec{\bigtriangledown} \cdot (1+\alpha (r)) \vec{%
\bigtriangledown} + \vec{L} \cdot \beta (r) \vec{L} +u(r) + W(r)
\label{eq:PS}
\end{equation}
with the parametrization \cite{Le} is used. The local ionic contribution is
carried in $u(r)$ and non-local effects lead to the effective mass $%
m^*(r)=m_e/(1+\alpha (r))$ and the orbital term $\propto\beta(r)$. Coulomb
and exchange-correlation potentials are represented by $W(r)$.

Results of our studies are exhibited in Figs. 2-4 and Table 1. In Fig. 2
the orbital M2 strength, $B(M2)=|<i|{\hat{F}}(M2)|0>|^2$, is
presented for light, medium and heavy spherical Na and K
clusters ($i$ represents $(1p-1h)_{2^-}$ states). Only the orbital part
of the M2 transition operator is used. Fig. 2 shows that dominant
twist strength is concentrated in the one 1p-1h peak with the lowest energy.
This relation persists independent of $N$ because both the twist mode
and other relevant 1p-1h excitations follow a trend
$\propto N_e^{-1/3}$ (see also Fig. 4).
The degree of concentration, however, changes with $N$. The twist peak
exhausts $100\%$, $80\%$, and $60\%$ of the total M2 strength in
$Na_9^{+}$, $Na_{93}^{+}$, and  $Na_{1315}^{+}$, respectively. It
corresponds to the nodeless branch of the twist mode (see Fig. 1,
left side). The nature of the peak is clarified in Fig. 3. It
represents $n,l \to n,l+1$ transition between single-particle levels with
the node number $n=1$ and {\it maximal} orbital moments $l$. The levels
belong to the last
occupied and first empty shells. As a result, the twist can serve as a
valuable source of information about i) single-particle levels with {\it %
maximal} orbital momenta near the Fermi surface and ii) the energy gap, $%
\Delta E_{sh}$, between the Fermi and next empty shells.

In large clusters weaker peaks also contribute to the twist resonance. As a
rule, they represent $n,l \to n,l+1$ transitions with $n=2,3,...$ and lower
orbital moments. Following our analysis of the velocity field, these peaks
also mainly contribute to the nodeless twist branch (Fig. 1,
left side). Their contribution grows with increasing cluster size. They come
energetically closer to the dominant peak (see Fig. 2) such that
they may not be easily distinguishable experimentally.
Other twist branches (see, e.g., the right side of Fig. 1
) carry only a small fraction of the total strength and lie
at higher energies.

Fig. 2 demonstrates similarity of the twist in Na and K
clusters. Much similar pattern appears for Li. However, there are also
distinctive differences in twist energies and strength. These values are
compared in Fig. 4. K, Na and Li are distinguished by
Wigner-Seitz radii (in atomic units, $r_s($K$)=5$, $r_s($Na$)=3.96$ and $%
r_s($Li$)=3.25$, respectively) and by different influence of the ionic
structure. Fig. 4 shows the following trends: i) The denser the
metal, the higher the twist energy. This can be explained by the fact that
the smaller the $r_s$, the deeper the corresponding single-particle
potential \cite{RPAclust}. In Li clusters the potential is most deep and,
therefore, has most large energy gap $\Delta E_{sh}$ between neighboring
quantum shells. Being close to $\Delta E_{sh}$, the twist energy should
increase from K to Li. ii) In all the size region the twist mode resides
far below the Mie dipole plasmon and, at the same time, it stays still far
above typical ionic vibrations $(\omega \simeq 50$ meV), which both helps for
an experimental discrimination. The heavier the cluster, the smaller the
energy gap $\Delta E_{sh}$ and, therefore, the twist energy. Fig. 4
 demonstrates the energy fall $\propto N_e^{-1/3}$. It is
worth noting that such dependence of twist energy takes place in both atomic
clusters and nuclei. iii) The denser the metal, the smaller the resonance
strength. As is shown below, the energy-weighted sum rule $S_l$
keeps the same value for K, Na and Li clusters of a given size (if one
neglects the ionic structure effects). So, the increase in the excitation
energy has to result in the decrease of the strength. In Li clusters the
strength is additionally suppressed due to the effective mass, $%
m^{*}/m_e\sim 1.2$. iv)
$B(M2)$ grows as $N_e^2$ (see upper part of Fig. 4).

All the trends discussed above are supported by the elastodynamical results (%
\ref{eq:elast}). However, the quantitative agreement is less perfect. Eqs. (%
\ref{eq:elast}) considerably overestimate both twist energy and strength (up
to $80\%$ and $30\%$ in light and heavy clusters, respectively). That is not
so surprising because we find that the twist mode is dominated by the shell
structure while the collective model averages over shells.

The attractive feature of the orbital M2 mode is that its total
strength can be estimated in a simple fashion by thwe energy-weighted
sum rule as
\begin{eqnarray}\label{eq:ewsr}
&&S=\sum_i\omega _i|<i|{\hat{F}}(M2)|0>|^2  \\
&\simeq &\frac 12\sum_{\mu =-2}^2\langle 0|[{\hat{F}}(M2\mu ),[H_0,{\hat{F}}%
(M2\mu )]]|0\rangle =S_s+S_l   \nonumber
\end{eqnarray}
where
\begin{equation}
S_s=\frac{75\hbar ^2}{2\pi m_e}\sum_{n_il_i}^{occ}(2l_i+1)\int (1+\alpha
+8r^2\beta )\rho _{n_il_i}dr,  \label{eq:ewsr_s}
\end{equation}
$$
S_l=\frac{25\hbar ^2}{3\pi m_e}\sum_{n_il_i}^{occ}(2l_i^3+3l_i^2+l_i)\int
(1+\alpha +\frac 85r^2\beta )\rho _{n_il_i}dr  \label{eq:ewsr_o}
$$

\noindent
are contribution of the spin and orbital parts of the transition operator ${%
\hat{F}}(M2)$. In Eq. (\ref{eq:ewsr}) the sum runs over all  $%
(1p-1h)_{2^{-}}$ states. In Eqs. (\ref{eq:ewsr_s}), $%
\alpha (r)$ and $\beta (r)$ are the functions responsible for the nonlocal
effects in the pseudo-Hamiltonian (\ref{eq:PS}), $\rho
_{n_il_i}=(rR_{n_il_i}(r))^2$, $R_{nl}(r)$ is radial wave function of the
single-particle state $nl$.
In Eqs.(\ref{eq:ewsr_s}), the sum runs over
all occupied single-particle levels. The spin-orbit coupling is neglected in
the calculations, but it would anyway not contribute directly to $S_l$.
Without the spin-orbit coupling the spin-dipole residual two-body
interaction also does not contribute to $S_l$.
The calculations show that even for Li clusters one may safely neglect the
spin-orbit correction $\beta$ because it contributes less than $0.1 \%$. The
influence of the  correction $\alpha$ is strong in Li: we
have $m^*_e/m_e \sim 1.2$\cite{EPJ} and so this correction decreases $S_l$
by a factor of $\sim 4/5$.

If one neglects the non-local corrections (which is justified for Na and
K  clusters\cite{PH,EPJ}), then the radial integrals in Eqs.(\ref{eq:ewsr_s}%
) are just 1 and both $S_s$ and $S_l$ become very simple
and even {\it model independent} (the similar sum rules have been derived
for the twist in atomic nuclei\cite{Schw} and spin-dipole excitations in
clusters\cite{SpinSR}). This is evident for $S_s$ where $%
2\sum_{n_il_i}^{occ}(2l_i+1)=N_e$. As for $S_l$, one should mention that
different models predict, as a rule, the same sequence of occupied levels in
spherical clusters, at least for light and medium sizes. Finally, it is
worth to emphasize that: i) the calculation of $S_l$ is extremely simple: it
is enough to know the orbital moments $l$ of occupied single-particle
levels; ii) both $S_s$ and $S_l$ values are equal for clusters of a given
size but of different metals (K, Na), which agrees with Eqs. (\ref
{eq:elast}).

The twist part, $\mu _b\sqrt{80/27}rY_{10}l_z$, of the transition operator ${%
\hat{F}}(M2)$ gives exactly $4/9$ of the complete $S_l$ values. Following
Fig. 1, the twist represents the oscillations in equatorial planes. The
total orbital M2 resonance takes also into account the meridian oscillations.

Results presented in Table 1 show that the orbital M2 energy-weighted
strength {\it dominates} over the spin one already in clusters with $N_e=40$%
. Starting with $N_e=92$, the orbital contribution becomes overwhelming and,
already for $N_e=440$, demonstrates a huge value of $2\cdot 10^5\mu _b^2\AA
^2eV$. Since the long-wave M1 response, both spin and orbital, is forbidden
in spherical clusters (indeed, both spin and orbital 1p-1h matrix elements
of M1 transition are proportional to the radial integral $\int
R_{n_1l_1}(r)R_{n_2l_2}r^2dr=\delta _{n_1l_1,n_2l_2}$ which is zero in the
nondiagonal case due to orthonormalization condition), the twist starting
with medium sizes becomes the {\it strongest} multipole magnetic mode. This
fact emphasizes its fundamental character.

Our calculations indicate that twist mode cannot be detected in
photoabsorption spectra since it is masked by low-energy E1
excitations. These excitations are much weaker than the dipole
plasmon but, nevertheless, strong enough to mask the twist. The
inelastic scattering of polarized optical photons (resonant Raman
scattering) seems to be more appropriate to observe the twist,
though any conclusions about perspectives of these reactions
still requires a careful analysis of the competition between E1,
E2 and M2 modes. The Raman scattering can separate electric and
magnetic modes due to the polarization selection rules. Clusters
with about $10^4$ atoms seem to be optimal. In such clusters the
twist strength reaches impressive values and, at the same time,
is strongly concentrated at very narrow low-energy interval
(which remains still well separated from ionic vibrations). Our
estimations show that in heavy clusters, in spite of a dense
spectrum, the twist exists and carries the spectroscopic
information mentioned above.

In summary, the M2 orbital resonance and its important part, the twist mode,
have been investigated in spherical alkali metal clusters. The macroscopic
treatment of the twist exhibits this mode as a general feature of any finite
3-dimensional Fermi system and provides a pertinent description of the basic
trends. However, it is not delivering a quantitative agreement. The
microscopic treatment, including the novel energy-weighted sum rules,
clarifies the main properties of twist 1p-1h M2 response. The resonance is
mainly exhausted, first of all in clusters of a moderate size, by one 1p-1h
M2 transition connecting the Fermi and next empty quantum shell, namely,
their levels with maximal orbital moments. As a result, the twist can
provide a valuable information about the single-particle scheme.
Twist energy and strength
evolve with a cluster size as $\omega \sim N_e^{-1/3}$ and $B(M2)\sim N_e^2$%
. In heavy clusters an impressive M2 strength can be reached. The twist {\it %
dominates} in the low-energy region over its spin-dipole counterpart already
in clusters of a moderate size and finally becomes the strongest magnetic
multipole mode. We hope that the fundamental significance of the twist for
orbital magnetism in spherical clusters will encourage experimentalists to
look for proper ways for its observation.

\vspace{0.2cm} {\bf Acknowledgments.} We thank E. Duval and N. Lo Iudice for
useful discussions and Ll. Serra for presenting single-particle schemes for $%
Li$ clusters. The work was partly supported (V.O.N.) by RFBR grant N
00-02-17194.

\newpage

{\bf FIGURE CAPTIONS.}

{\bf Figure 1}:
Nodeless (left) and one-node (right) branches of twist mode
\protect\cite{HE}.
\vspace{5mm}

{\bf Figure 2}:
The distribution of $M2$-strength in spherical Na (left) and K
(right) clusters of different sizes, as indicated.
\vspace{5mm}

{\bf Figure 3}:
Single-particle levels and M2 1p-1h transitions in Na$^+_{9}$ and
Na$^+_{93}$. The Fermi levels are marked by the double line. For the main
transitions (bold arrows), the contributions to the complete strength $B(M2)$
are given.
\vspace{5mm}

{\bf Figure 4}:
The strength normalized by $N_e^2$ (upper panel) and averaged
energy (lower panel) of the twist resonance in K, Na and Li clusters.
The trends $B(M2)\propto N_e^2$ and ${\bar{\omega}}\propto N_e^{-1/3}$ are
distinctive.

\begin{table}[tbp]
\caption{Orbital energy-weighted sum rule $S_l$ (in units $\mu_b^2 \AA^2 eV$%
) and ratio $R=S_l/S_s$ for Na and Li spherical clusters in the energy
interval 0-6 eV ($95-99\%$ of the sum rules are exhausted in the interval
0-2 eV). Values $R$ are equal for K, Na and Li clusters of the same
size. }
\label{tab:sr}\vspace{0.4cm} 
\par
\begin{center}
\begin{tabular}{|c|c|c|c|}
$N_e$ & \multicolumn{2}{|c|}{$S_l$} & $R$ \\ \cline{2-3}
& K, Na & Li & K, Na, Li \\ \hline
8 & 1.21$\cdot 10^2$ & 1.08$\cdot 10^2$ & 0.33 \\
20 & 7.28$\cdot 10^2$ & 6.42$\cdot 10^2$ & 0.80 \\
40 & 2.55$\cdot 10^3$ & 2.23$\cdot 10^3$ & 1.40 \\
92 & 1.35$\cdot 10^4$ & 1.15$\cdot 10^4$ & 3.21 \\
440 & 2.01$\cdot 10^5$ & 1.67$\cdot 10^5$ & 9.72
\end{tabular}
\end{center}
\end{table}

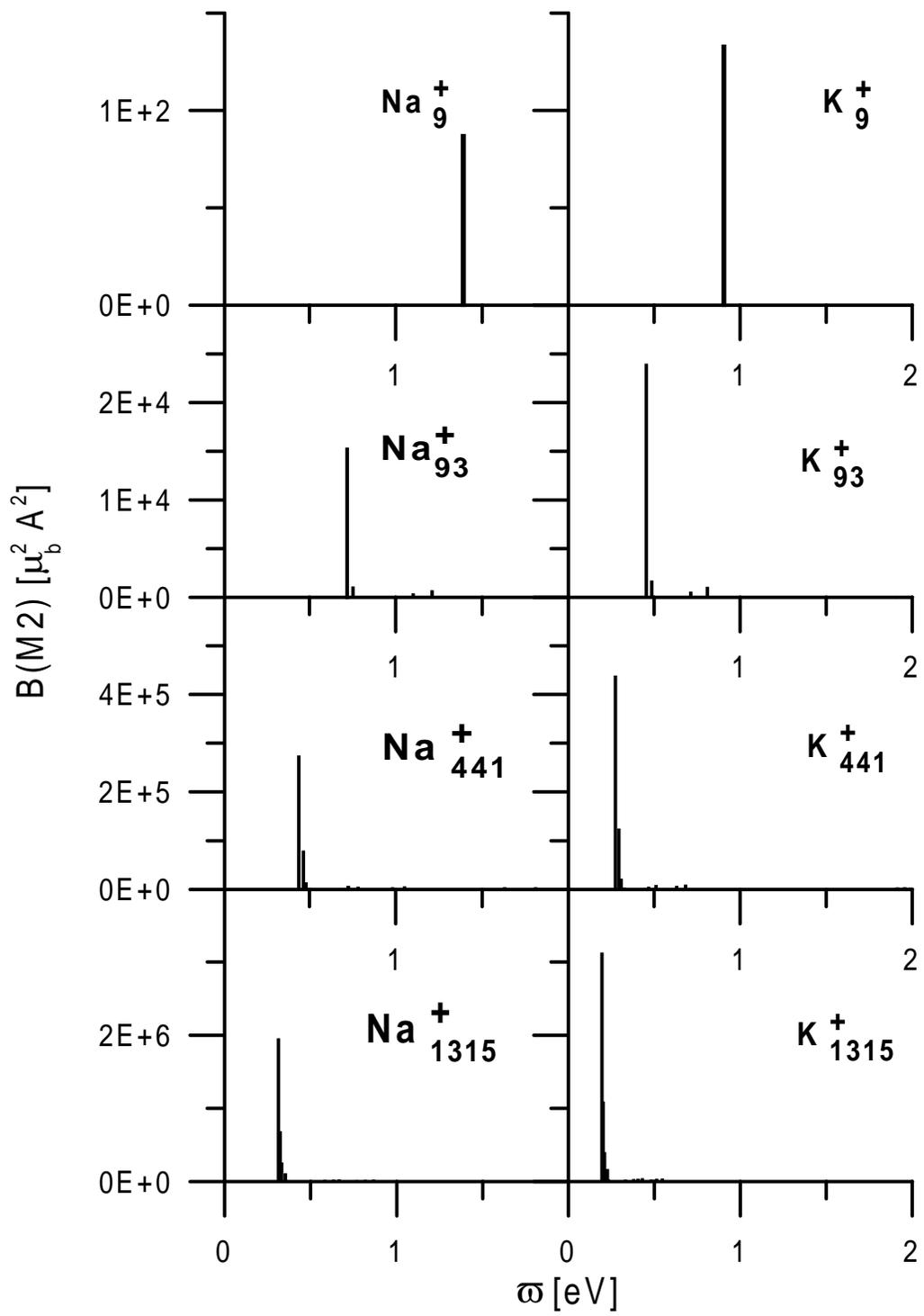
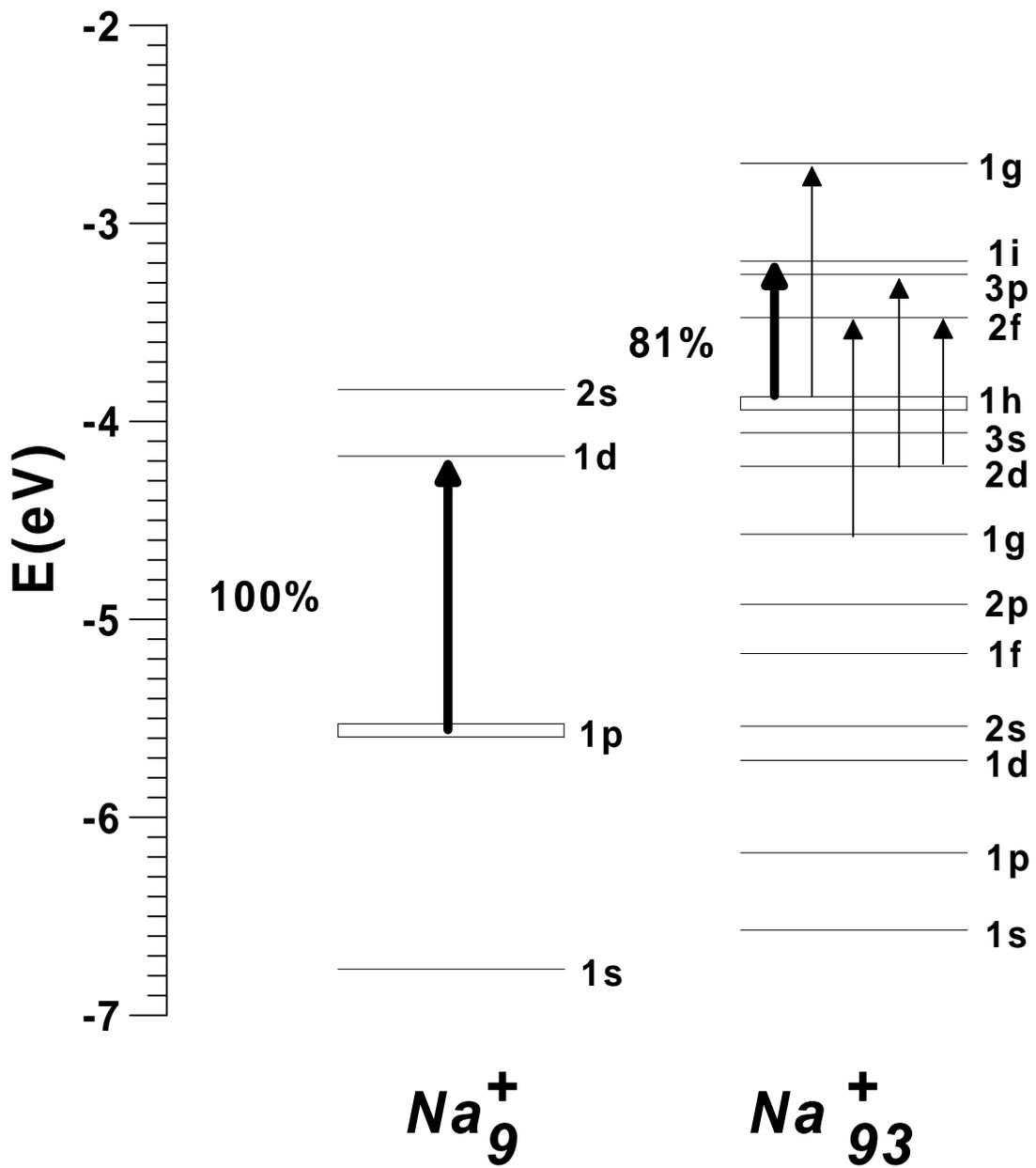
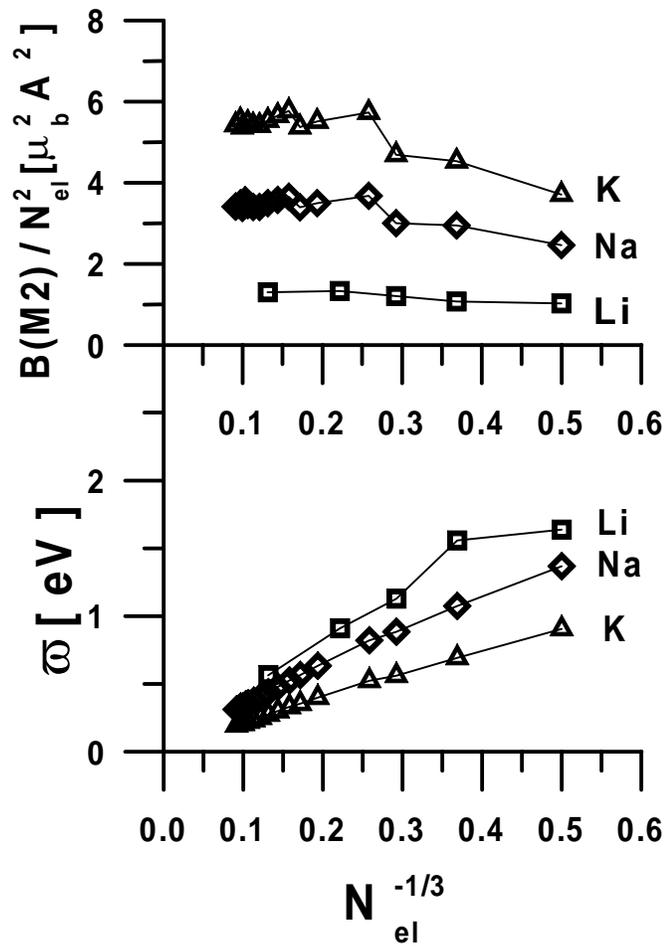
\end{document}